\documentclass[aps,preprint,showpacs,superscriptaddress]{revtex4}

\usepackage{graphicx}
\usepackage{amsmath}
\usepackage{bm}

\newcommand{\be}{\begin{equation}}
\newcommand{\ee}{\end{equation}}

\begin{document}

\title
{Phonon drag of electrons at high temperatures}

\author{V.A. Vdovenkov}

\affiliation{Moscow State Institute of Radioengineering,
 Electronics and Automation (technical university)\\
Vernadsky~ave. 78,~117454~Moscow,~Russia}

\begin{abstract}
Temperature-dependent thermoelectric power (TEP) in semiconductor
crystals and crystal structures containing an electron-vibrational
centers (EVC) was investigated in this article. The TEP contain
narrow pius at Debye temperatures for different phonons. In thin
epitaxial layers on substrates such pius exist at Debye
temperatures of substrates phonons. These pius existence imposable
to explain on basis of known TEP  theory but it may be explained
by the phonon drag of electrons (PDE) effect. In this connection
there is the necessity to change traditional point of view on the
PDE effect existence only at low temperatures and expand the PDE
theory on case of strong electron-phonon coupling provided by EVC
at high temperatures.
\end{abstract}



\maketitle

\section{Introduction}
Thermoelectric power phenomenon  was found out in 1821.  This
phenomenon arises in current-conducting materials at presence of
temperature gradient. This physical phenomenon is caused by
diffusion of mobile charges carriers in volume of a material under
temperature gradient and is considered to be proportional to the
temperature gradient. Factor of proportionality consist of two
components which take into account the contributions of electrons
($\alpha_{n}$) and holes ($\alpha_{p}$) \cite{Seeg73} :
\begin{equation}\label{E1}
\alpha = \alpha_{n} + \alpha_{h}.
\end{equation}
If relaxation time of wave pulses for electrons and holes depend
on energy (E) as $E^{r}$ then in nondegenerated semiconductor
\begin{equation}\label{E2}
\alpha_{n} = - \frac{k}{|e|}(r + \frac{5}{2} -
\frac{F_{e}}{kT}),~~~~~~~~~~ \alpha_{h} = + \frac{k}{|e|}(r +
\frac{5}{2} - \frac{F_{h}}{kT}),
\end{equation}
where  $k$ - Boltzman constant, e - electron charge,  r -
dissipation parameter, $F_{e}$ and $F_{h}$ - Fermi levels for
electrons and holes, T - temperature. In some semiconductor
materials at low temperatures the experimental magnitude of TEP
exceed the magnitude which was predicted on basis of the TEP
diffusion theory. This disagreement between theory and experiment
was explained by another physical phenomenon - phonon drag of
electrons (PDE). The PDE represents the additional contribution in
thermoelectric power. The PDE component is not connected with
diffusion of mobile charges carriers but is caused by their moving
under phonons flow action due to electron-phonon coupling.

For the first time L. Gurevich has predicted the PDE existence  in
metals and has specified the opportunity of the phenomenon
existence in semiconductors \cite{Gur45}. G. Pikus investigated
this phenomenon theoretically in semiconductors and in 1951 has
deduced the formula for PDE which arise as a result of interaction
of incoherent phonons flow with mobile charge carriers. He has
shown that the magnitude of the PDE depend on the relation between
relaxation times for wave pulses of phonons and electrons and has
come to a conclusion about insignificant magnitude of the effect.
However it was shown experimentally that the PDE in semiconductors
can many times over exceed the magnitude which predicted by
diffusion theory of thermoelectric power \cite{Fred53} -
\cite{Geba54}. It was established that the PDE is caused by flow
of those phonons which interact with electrons \cite{Fred53,
Herr53, Herr58}. Or else, the PDE arises at presence of coupling
between electrons and phonons.

The PDE component of TEP which take into account phonons flow
interaction with mobile electrons may be described by coefficient
\begin{equation}\label{E3}
\alpha_{ph} =
(\frac{k}{e})\frac{m^{*}v^{2}}{kT}\cdot\frac{<\tau_{ph}>}{<\tau>}
\end{equation}
where $<\tau_{ph}>$ - middle time for phonons wave pulse
relaxation, $<\tau>$ - middle time for electrons wave pulse
relaxation, $m^{*}$ - fictitious mass and $v$ - speed of movable
charge carriers \cite{Herr58}. In accordance to counting the
long-wave acoustical phonons participation  in the PDE effect the
coefficient $\alpha_{ph}$ depend from temperature as $T^{-7/2}$
and decrease at heating.

The PDE effect was observed at temperatures above 70 K. Ones came
to a conclusion that given effect is impossible to detect at
T~$>$~70~K. Absence of the given effect at T $>$ 70~K was
explained by weak electron-phonon coupling. Therefore the
semi-conductor materials containing electron-vibrational centers
(EVC) are attractive for supervision of the PDE effect
\cite{arX99} because inherent oscillations of such centers are
capable to couple with electrons, phonons and hence to supply
strong coupling between electrons and phonons.

The purpose of the given work is to research the phonon drag of
electrons in crystals and crystal structures containing an
electron-vibrational  centers.

\section{Experiments, results and discussion}
\subsection{Description of experiments}
Industrial polished semiconductor plates by thickness 200 microns
were used in our experiments. The semiconductor samples contained
local electron-vibrational centers  with representative strong
electron-phonon coupling.

GaP samples with impurities  of aluminum GaP(Al) or sulphur GaP(S)
were investigated. Impurity concentrations of Al or S was equal to
about $5 \cdot 10^{15}$cm$^{-3}$. Such impurities were chosen
because atoms Al and S have masses vastly exceed  mass of atom Ga.
This promote emergence of electron-vibrational centers. Before
measurements the samples were subjected to heating in vacuum
within 5 minutes at temperature T = 600~K and cooled up to room
temperature within 0.2 minutes to activate EVC.

Samples of Si, InSb, InAs were irradiated by fast electrons
integer flow $\sim10^{18}~cm^{-2}$ with energy 1 MeV to introduce
EVC. Silicon samples with impurities of phosphorus
($\approx~5\cdot 10^{15}$cm$^{-3}$) and oxygen
($\approx~10^{18}$cm$^{-3}$) : Si(P,O) were used. Concentration of
oxygen impurity in Si was determined on  basis of the data about
distinctive for oxygen impurity optical absorption in spectral
band near 9 mkm \cite{Kais56}. The main type of EVC in silicon
samples was A-center which represent association of impurity
oxygen atom with vacancy. In accordance to results of IR
absorption spectra measurements the concentration of A-centers was
equal to $\approx 10^{15} cm^{-3}$ and the Huang-Rhyce constant
\cite{Hua50} for electron-phonon interaction $S \simeq 5$. Single
crystals of InSb and InAs were investigated. Initial concentration
of mobile electrons (before irradiation by fast electrons) at room
temperature was equal $1.68\cdot10^{14} cm^{-3}$ for InSb and
$6\cdot10^{12} cm^{-3}$ for InAs.  Epitaxial layers of InSb and
InAs with thickness up to 20 mkm on semiinsulator GaAs substrates
with thickness 200 mkm were investigated else. Concentration of
mobile electrons in initial InSb layer (before irradiation by fast
electrons) was equal to $3\cdot10^{16} cm^{-3}$. Concentration of
mobile electrons in initial InAs layer  was equal to
$3\cdot10^{16} cm^{-3}$, but after the irradiation it was near to
inherent concentration. Supposedly EVC in the samples were formed
by impurity oxygen atoms which usually are contained in
significant concentrations in these semiconductor materials.

We investigated the carbon nanotube films with  thickness
$\simeq$~0.1 mkm with surface resistance $\cong 10^{3}~Ohm \cdot
cm^{-2}$ on quartz substrate. These samples were fabricated by
dispersion of graphite single crystal by electron beam in vacuum
\cite{Kos92}. Carbon nanotube films where consisted from single
walled nanotubes. Diameter of nanotubes was about 10~nm. The
nanotubes were oriented along normal to substrate surface and form
two-dimensional regular structure. In accordance to investigated
IR reflectance spectra and  activation energies of resistivity the
EVC in these samples were formed by carbon atoms. Supposedly the
EVC existence is connected with variation of nanotube chirality.

The TEP were measured at temperatures  is higher 77~K where the
PDE existence was considered as impossible. Temperature difference
between electrical contacts to every sample was not more than 3K,
probable inaccuracy at temperature measurements was not more than
0.2~K and at voltage measurements it was not more than
$10^{-6}$~V.

\subsection{Experimental results}

\begin{figure}
\vspace*{-1cm}
\includegraphics[width=8cm]{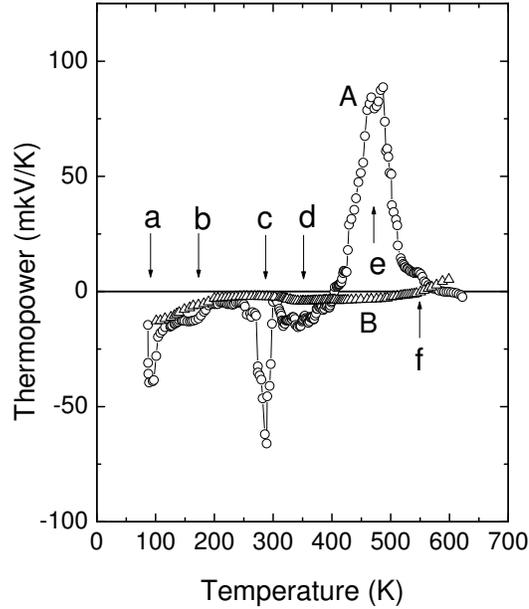}
\vspace*{-1cm} \caption {Experimental temperature dependency of
thermoelectric power in GaP with impurity of Sulphur (A) and
calculated temperature dependency of thermoelectric power in GaP
with close to inherent conductivity (B).} \label{fig1}
\end{figure}

Typical temperature dependency of thermoelectric power for GaP(S)
samples is shown in Fig.~\ref{fig1} by curve~A.  Curve~A keeps
pius which marked by arrows and by letters. Temperature dependency
of the TEP for GaP(Al) is like the curve~A. Curve~B in
Fig.~\ref{fig1} is the TEP temperature dependency which was
calculated by using Eqs.~(\ref{E1}, \ref{E2}) for GaP sample with
close to inherent conductivity.

\begin{figure}
\vspace*{0cm}
\includegraphics[width=8cm]{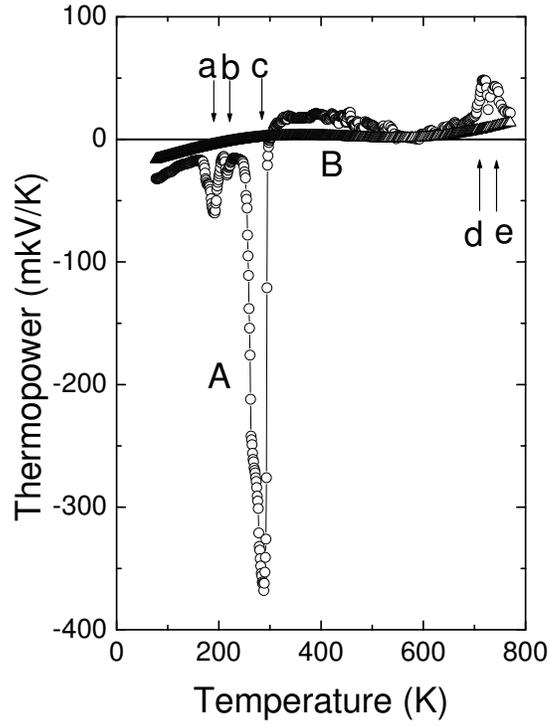}
\vspace*{0cm} \caption {Temperature dependency of thermoelectric
power: experimental for Si containing A-centers (A) and calculated
for Si with close to inherent conductivity (B).} \label{fig2}
\end{figure}

Typical temperature dependency of thermoelectric power  for
Si(P,O) samples is shown in Fig.~\ref{fig2} by curve~A. This curve
has pius specified by arrows and by letters. Curve B on
Fig.~\ref{fig2} is temperature dependency of the TEP for Si single
crystal with close to inherent conductivity which was calculated
by using Eqs.~(\ref{E1}, \ref{E2}).

Temperature dependencies of the TEP for irradiated by fast
electrons single crystals of InSb and InAs are shown in
Fig.~\ref{fig3}. Characteristic temperature dependency of
thermoelectric power for irradiated epitaxial layer of InSb on
semi-insulating GaAs substrate (InSb/GaAs) is represented in
Fig.~\ref{fig4}. Some pius on the curve are pointed out by arrows.
\begin{figure}
\vspace*{0cm}
\includegraphics[width=10cm]{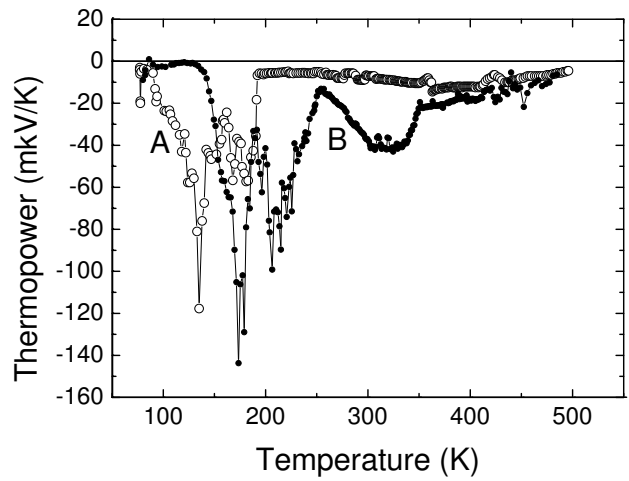}
\vspace*{0cm} \caption {Temperature dependency of thermoelectric
power in irradiated by electrons single crystals InSb (A) and InAs
(B).} \label{fig3} \label{fig3}
\end{figure}
\begin{figure}
\vspace*{-2cm}
\includegraphics[width=8cm]{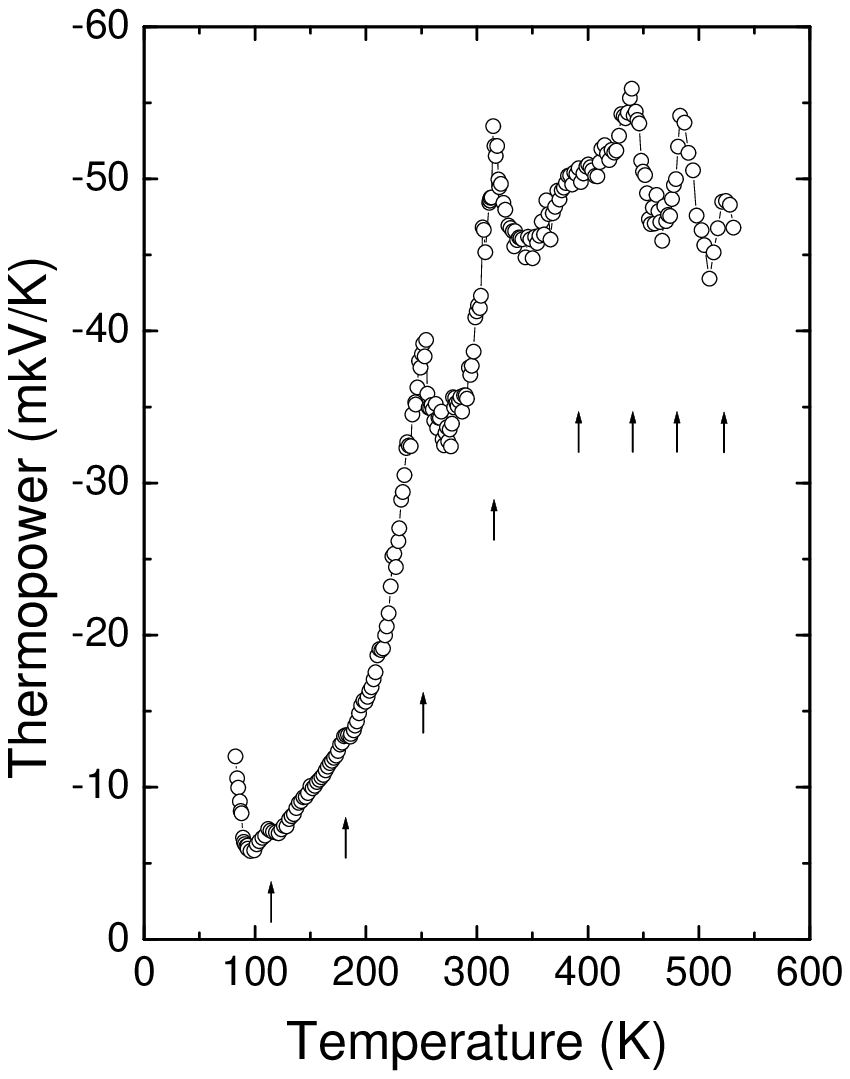}
\vspace*{-1cm} \caption {Temperature dependencies of
thermoelectric power in irradiated by fast electrons InSb
epitaxial layer on semi-insulating GaAS substrate. Some pius of
the dependencies are pointed out by arrows.} \label{fig4}
\end{figure}
Characteristic temperature dependencies of thermoelectric power
for irradiated by fast electrons epitaxial layer of InAs on
semi-insulating GaAs substrate (InAs/GaAs) are represented in
Fig.~\ref{fig5}.

\begin{figure}
\vspace*{-1cm}
\includegraphics[width=8cm]{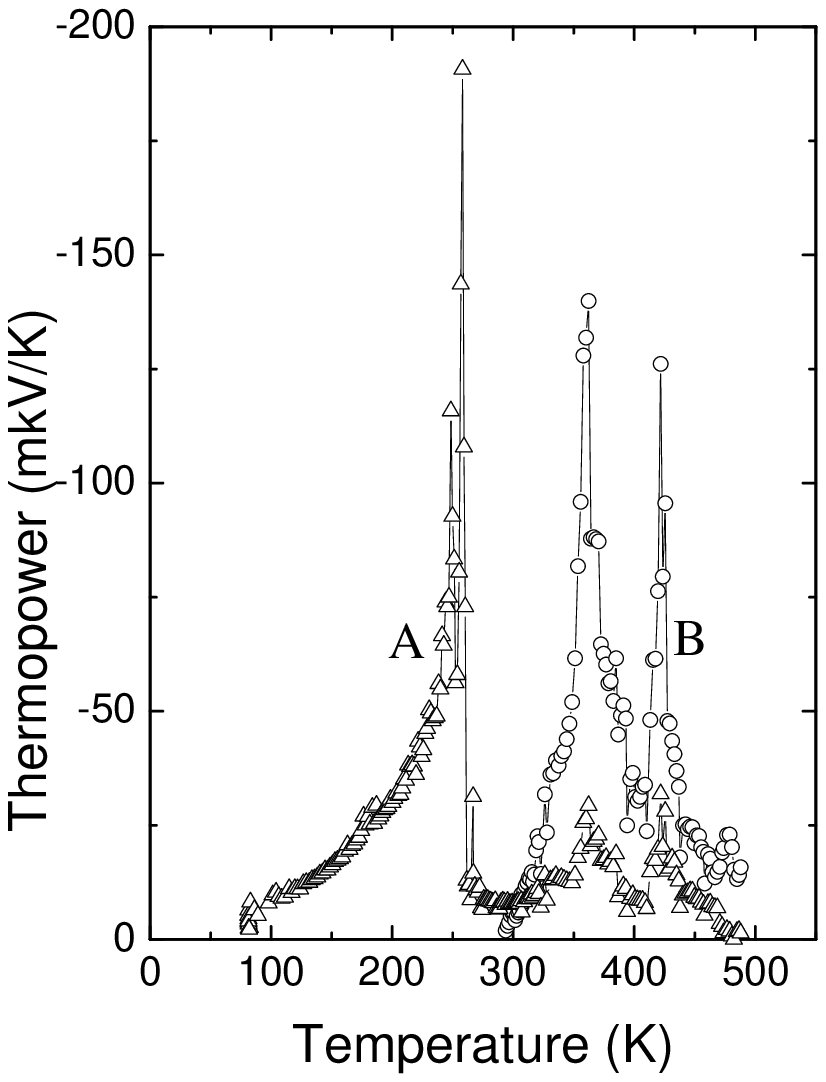}
\vspace*{-1cm} \caption {Temperature dependencies of
thermoelectric power in irradiated InAs epitaxial layer on
semi-insulating GaAS substrate. Curve A was measured at heating
but curve B was measured at cooling of the sample.} \label{fig5}
\end{figure}
\begin{figure}
\vspace*{-1cm}
\includegraphics[width=8cm]{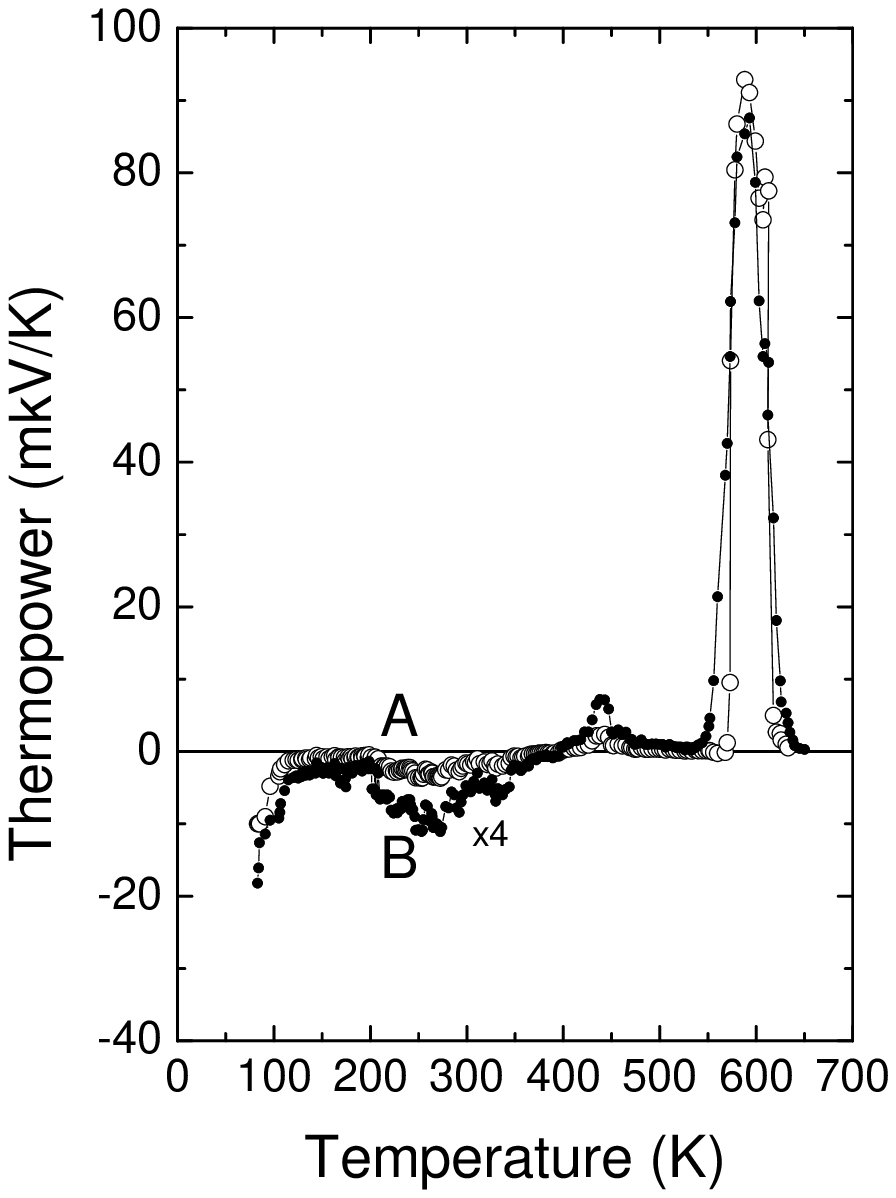}
\vspace*{-1cm} \caption {Temperature dependency of thermoelectric
power in carbon nanotube film on quartz substrate (A) and in
silicon epitaxial layer on quartz substrate (B). Curve B is
increased in 4 times in relation to valid curve.} \label{fig6}
\end{figure}

Typical temperature dependencies of thermoelectric power for
carbon nanotube film is shown in Fig.~\ref{fig6} by curve A. The
TEP for silicon epitaxial layer by thickness 14 mkm on quartz
substrate with impurity of Phosphorus ($\simeq 10^{14} cm^{-3}$)
is shown in Fig.~\ref{fig6} by curve B which is increased in 4
times in relation to valid curve.

\subsection{Discussion}

Curve A in Fig.~\ref{fig1} contain pius whose polarity agree with
the diffusion TEP polarity. Pius~a, b, c, f are situated at Debye
temperatures of  phonons in GaP~\cite{Mar64}: 95~K (TA,~8.2~meV);
168~K (TA,~14.25~meV); 288~K (LA,~24.42~meV); 542~K
(LO,~44.75~meV). Broad pius~d~($\approx345K$) and
e~($\approx475K$) may to be explained by combinations of phonons:
(TA + TA,~28.6~meV) and (TA + LA,~38.67~meV). Curve B in
Fig.~\ref{fig1} is the TEP which was calculated for GaP with close
to inherent conductivity and without an account of interaction
between electrons and phonons flow. This curve is smooth and does
not contain any pius. Therefore located at Deby temperatures of
phonons the narrow PDE pius in GaP(S) and GaP(Al) definitely may
be connected with the EVC which are formed by impurity atoms of
Sulfur or Aluminum.

Investigations of resistivity temperature dependencies,  infra-red
reflection spectra and the TEP temperature dependencies in GaP(Al)
and GaP(S) show that impurity atoms of Al and S form the EVC in
GaP due to which nuclei oscillations (inherent, I-oscillations) of
Al and S actively interact with phonons and electrons and supply
strong electron-phonon coupling. The I-oscillations can spread in
crystals, interact with electrons and give rise to waves of
I-oscillations and electrical currents. This is confirmed by
particularities of the TEP.
 In connection with the given results it is possible to
consider that I-oscillations of the EVC,  which  are formed by
impurity atoms Al and S, are the reason of the narrow TEP peaks at
Deby temperatures of phonons. It is possible to explain pius a, b,
c, d by phonon drag of electrons and pius e, f by phonon drag of
holes at Deby temperatures of the same phonons which effectively
interact with EVC and supply strong electron-phonon coupling.

Curve A in Fig.~\ref{fig2} contain pius~a, b, c which are situated
at Debye temperatures of known characteristic acoustic phonons in
Si whose wave vectors directed along certain directions in
Brillouin zone~\cite{Aub63}: $<$111$>$,~200.4~K (16.7~meV);
$<$110$>$,~214.8~K (17.9~meV); $<$100$>$,~252~K (21.0~meV). Curve
B on Fig.~\ref{fig2} is the TEP temperature dependence which was
calculated for Si with close to inherent conductivity without any
account of interaction between electrons and phonons flow.
Comparison of curves A and B in Fig.~\ref{fig2} allows to connect
the pius~a, b, c in curve A with presence of A - centers and to
explain them  by phonon drag of electrons. Pius d and e we connect
with TO phonons drag of holes in~Si.

Represented in Fig.~\ref{fig3} the TEP temperature dependencies
contain some narrow pius. The temperatures of most significant TEP
pius for InSb are submitted in Table~I together with Deby
temperatures, energies and types of determined optically suitable
phonons in InSb single crystal \cite{Mar64, Lyd41}. Taking into
account the data from Table I the  TEP peaks in InSb may be
connected with phonons participation.
\begin{table*}[t]
\begin{center}
\caption{Temperatures of the TEP pius, Deby temperatures
($T_{D}$), energies ($E_{p}$) and types (Type) of suitable phonons
in InSb single crystal.} \vspace{0.5cm}
\begin{tabular}{|c|c|c|c|c|c|c|c|c|c|c|}
\hline \hline
\ T~(K)& 134 & 168 & 182 & 191 & 273 & 293 & 430 \\
\hline
\ $T_{D}$~(K) & 124 & 170 & 170 & 170 & 266 & 280 & 446 \\
\hline
\ $E_{p}~(meV)$ & 5.3 & 14.6 & 14.6 & 14.6 & 22.9 & 24.4 & 16.2 ~\\
\hline
\ ~~~~Type~~~~&~~~2TA~~~&~~~~LA~~~~&~~~~LA~~~~&~~~~LA~~~~&~~~~LO($\Gamma$)~~~~&~~~~LO($\Gamma$)~~~~&~~~2TO~~~\\
\hline \hline
\end{tabular}
\end{center}
\end{table*}
Identical data concerning InAs sample with data about suitable
phonons in InAs single crystal \cite{Lor65} are submitted in Table
II. Contained in the Table II data allow to connect the TEP pius
in InAs with participation of phonons. It is visible from Table II
that the most intensive PDE pius at temperatures 173~K and 178~K
have not suitable phonons. These pius differ from the PDE pius at
temperatures 208~K and 212~K on $\simeq 34 K$ and may be explained
by participation in the PDE effect of acoustical phonons with
$T_{D} \simeq 34~K ~(2.9 meV)$. It may be possible because such
acoustical phonons can influence on EVC inherent oscillations,
make them coherent, enable to arise coherent areas, heat
superconductivity and hyperconductivity in different
semiconductors \cite{arX00}.
\begin{table*}[t]
\begin{center}
\caption{Temperatures of the TEP pius, Debye temperatures
($T_{D}$), energies ($E_{p}$) and types (Type) of suitable phonons
in  InAs single crystal.} \vspace{0.5cm}
\begin{tabular}{|c|c|c|c|c|c|c|c|c|c|c|}
\hline \hline
\ T~(K)& 159 & 173 & 178 & 196 & 208 & 212 & 310 & 320 & 424\\
\hline
\ $T_{D}$~(K) & 102 & - & - & 206 & 206 & 206 & 319 & 319& 411 \\
\hline
\ $E_{p}~(meV)$ & 8.8 & - & - & 17.7 & 17.7 & 17.7 & 27.5 & 27.5 & 17.7 \\
\hline
\ Type & ~~~TA~~~&~~~~-~~~~&~~~~-~~~~&~~~LA~~~&~~~LA~~~&~~~LA~~~&~~~$TO_{1}$~~~&~~~$TO_{1}$~~~&~~~2LA~~~\\
\hline \hline
\end{tabular}
\end{center}
\end{table*}

One can see from  Fig.~\ref{fig4} that temperature dependence of
the TEP for InSb/GaAs contain some pius at different temperatures
though according to the TEP theory similar curves for InSb and
GaAs can not contain of such narrow pius.  Temperature of these
pius are inserted in Table III. Debye temperatures  of suitable
phonons energies and types of the phonons for single crystals InSb
\cite{Lyd41, Mar64} and GaAs \cite{Coc61} are inserted in the
Table III also. It is visible from Table III that the temperatures
of some pius TEP cannot confidently be identified with phonons in
InSb and their greater number is possible to explain only by
combination of such phonons. Examination of contained in Table III
data enable with the more probability to refer  the TEP pius in
InSb/GaAs to participation of GaAs substrate phonons.
\begin{table*}[t]
\begin{center}
\caption{Temperatures of the TEP pius in InSb epitaxial layer on
semiinsulator GaAs substrate (EL), Debye temperatures ($T_{D}$) ,
energies ($E_{p}$) and types (Type) of suitable phonons in InSb
and GaAs.} \vspace{0.5cm}
\begin{tabular}{|c|c|c|c|c|c|c|}
\hline \hline
\ ~EL~ & \multicolumn{3}{c|}{InSb} &  \multicolumn{3}{c|}{GaAs}  \\
\hline
\ T &~~$T_{D}$~~& $E_{p}$ & Type &~~$ T_{D}$~~& $E_{p}$ & Type  \\
\ (K) & (K) & (meV) & ~ & (K) & (meV) & ~ \\
\hline
\ 112 & - & - & -& 104 & 9.0 & TA \\
\hline
\ 181 & - & - & - & 182 & 15.7 & TA \\
\hline
\ 253 & 257 & 22.2 & TO & 258 & 22.2 & LA \\
\hline
\ 315 & 319 & - & TO + TA & 336 & 29.0 & LO \\
\hline
\ 383 & 393 & - & LO + LA & 393 & 33.8 & TO($\Gamma$) \\
\hline
\ 440 & 446 & 36.7 & 2LO & 425 & 36.7 & LO($\Gamma$) \\
\hline
\ 486 & 480 & - & TO + LO & 478 & - & TO + LO \\
\hline
\ ~~~~~523~~~~~&~~~~~532~~~~~&~~~~~23.3~~~~~~&~~~~2TO($\Gamma$)~~~~&~~~~~540~~~~~&~~~~~23.3~~~~~&~~~~~2LA~~~~~\\
\hline \hline
\end{tabular}
\end{center}
\end{table*}
\begin{table*}[t]
\begin{center}
\caption{Temperatures of the TEP pius in InAs epitaxial layer (EL)
on semiinsulator GaAs substrate, Debye temperatures ($T_{D}$),
energies ($E_{p}$) and types (Type) of suitable phonons in InAs
and GaAs.} \vspace{0.5cm}
\begin{tabular}{|c|c|c|c|c|c|c|}
\hline \hline
\ ~EL~ & \multicolumn{3}{c|}{InAs} & \multicolumn{3}{|c|}{GaAs} \\
\hline
\ T & $T_{D}$ & $E_{p}$ & Type &~$T_{D}~$ & $E_{p}$ & Type \\
\ (K) & (K) & (meV) & ~ & (K) & (meV) & ~ \\
\hline
\ 104  & 102 & 8.8 & TA & 104 & 9.0 & TA \\
\hline
\ 183 & - & - & - & 182 & 15.7 & TA \\
\hline
\ 248  & - & - & - & 248 & 21.4 & LA \\
\hline
\ 258  & - & - & - & 258 & 22.2 & LA \\
\hline
\ 267 & 282 & 24.3 & LO & 270 & 23.3 & LA \\
\hline
\ 326 & 319 & 27.5 & $TO_{1}$ & 336 & 29.0 & LO \\
\hline
\ 362 & 383 & - & LO + TA & 367 & 31.6 & TO \\
\hline
\ ~~~~422~~~~~&~~~~~410~~~~~&~~~~~17.7~~~~~&~~~~~2LA~~~~~&~~~~~425~~~~~&~~~~~36.7~~~~~&~~~~~LO($\Gamma$)~~~~~\\
\hline \hline
\end{tabular}
\end{center}
\end{table*}
The  temperatures of TEP pius visible in Fig.~\ref{fig5} for
InAs/GaAs are inserted in Table IV together with Debye
temperatures and energies of suitable phonons in  single crystals
InAs \cite{Lor65} and GaAs \cite{Coc61}. Examination of inserted
in Table IV data enable with high probability to connect the TEP
pius in InAs/GaAs with definite role of GaAs substrate phonons.

Represented in Fig.~\ref{fig6} temperature dependencies of the TEP
for carbon nanotube film and for silicon film on quatz substrate
are similar with each other in common features. This similarity of
curves A and B in Fig.~\ref {fig6} give definite basis to connect
the pius of the curves with participation of quartz substrate
phonons in phonon drag phenomenon.

\subsection{Conclusion}

Earlier the PDE in semiconductors was observed as narrow pius of
the TEP at temperatures below 70 K. Taking into account the new
data about the PDE pius located  close to Debye temperatures of
phonons the well known PDE effect at low temperatures in different
materials  may be explained as PDE pius caused by participation of
acoustical phonons with low Debye temperatures $(T_{D} < 70~K)$.
Absence of the PDE at T~$>$~70 K earlier was explained by
insufficiently strong coupling between electrons and phonons.
Nevertheless the temperature dependence of thermoelectric power in
ropes of carbon nanotubes supposedly was explained by the PDE
contribution at temperatures from 4.2~K to 300~K \cite{Hon98}. The
narrow thermoelectric power pius observable in semiconductor
materials and structures undoubtedly are connected with the phonon
drag of electrons (or holes). It is necessary to take into account
that the phonons frequencies can be changed or splitted under
interactions between EVC and appropriate Debye temperatures can be
changed. Therefore there is no basis to wait of strict concurrence
of the TEP pius temperatures with Debye temperatures for phonons.
The TEP pius in semiconductors single crystals are located near to
Debye temperatures of phonons but in epitaxial semiconductor
layers on semiinsulator substrates such pius are located near to
Debye temperatures of substrate phonons. The electron-phonon
coupling which supplied by electron-vibrational centers in
semiconductor crystals, in epitaxial layers and else in carbon
nanotube films on substrates probably is sufficiently strong for
the PDE effect realization even at high Debye temperatures.

The represented experimental data specify participation in PDE
effect not only long-wave acoustic phonons which are taken into
account in known PDE theory  but also  short-wave acoustic and
optical phonons in points of Brillouin zone with high density of
phonons frequencies. In connection with these data it is desirable
to expand the PDE theory on a case of strong electron-phonon
interaction provided by electron-vibrational centers at high
temperatures.

The investigation of the PDE pius  enable to study the phonons
types, their energies, Debye temperatures in containing
electron-vibrational centers semiconductor materials and
structures.

\begin{acknowledgments}
I thank Dr. Z. Ya. Kosakovskaya for carbon nanotube films on
substrates given for researches.
\end{acknowledgments}

\end{document}